\documentclass[aps,twocolumn,showpacs]{revtex4}
\usepackage{graphicx}
\def\bsigma{\mbox{\boldmath $\sigma$}}
\def\bmu{\mbox{\boldmath $\mu$}}
\def\bOmega{\mbox{\boldmath $\Omega$}}

\begin{document}
\title{Frequency dependence of induced spin polarization
and spin current in quantum wells}
\author{O. E. Raichev}
\email{raichev@isp.kiev.ua}
\affiliation{Institute of Semiconductor Physics,
National Academy of Sciences of Ukraine,
Prospekt Nauki 45, 03028, Kiev, Ukraine}
\date{\today}

\begin{abstract}

Dynamic response of two-dimensional electron systems with spin-orbit
interaction is studied theoretically, on the basis of quantum kinetic
equation taking into account elastic scattering of electrons. The spin
polarization and spin current induced by the applied electric field are
calculated for the whole class of electron systems described by ${\bf p}$-linear
spin-orbit Hamiltonians. The absence of non-equilibrium intrinsic static
spin currents is confirmed for these systems with arbitrary (non-parabolic)
electron energy spectrum. Relations between the spin polarization, spin
current, and electric current are established. The general results are
applied to the quantum wells grown in [001] and [110] crystallographic
directions, with both Rashba and Dresselhaus types of spin-orbit coupling.
It is shown that the existence of the fixed (momentum-independent)
precession axes in [001]-grown wells with equal Rashba and Dresselhaus
spin velocities or in symmetric [110]-grown wells leads to vanishing
spin polarizability at arbitrary frequency $\omega$ of the applied
electric field. This property is explained by the absence of
Dyakonov-Perel-Kachorovskii spin relaxation for the spins polarized
along these precession axes. As a result, a considerable frequency
dispersion of spin polarization at very low $\omega$ in the vicinity
of the fixed precession axes is predicted. Possible effects of extrinsic
spin-orbit coupling on the obtained results are discussed.

\end{abstract}


\maketitle

\section{Introduction}

The presence of spin-orbit interaction in solids provides a natural
way of manipulating spin states of electrons by purely electrical means,
without application of a magnetic field. This property is a subject
of interest for the novel and rapidly developing field of spintronics.$^1$
Some important manifestations of spin manipulation are the generation of
non-equilibrium spin polarization of electrons and excitation of spin
currents by a driving electric field which also leads to the usual
charge current. Though the problem of field-induced spin polarization
is an old one,$^{2,3,4,5}$ it has been recently set at the focus of
attention. The main reason for this is the appearance of experimental
works$^{6,7,8,9,10,11}$ which demonstrate the spin polarization generated
under the current flow both in bulk and two-dimensional (2D) semiconductor
layers with spin-orbit interaction. Another reason is the rapidly growing
interest to the related phenomenon, the intrinsic spin-Hall
effect, when the electric current in the presence of spin-orbit coupling
leads to a non-equilibrium spin current in perpendicular direction (note
that weak spin currents can exist even in equilibrium$^{12}$). After the
theoretical proposal$^{13}$ of the intrinsic spin-Hall effect based on the
Rashba model representing spin-orbit interaction for 2D electrons in quantum
wells,$^{14}$ it has been realized$^{15,16,17,18,19,20,21}$ that this
effect is absent in the infinite 2D system in the static (zero-frequency)
limit and exists at non-zero frequency or in finite-size samples.
The absence of the static intrinsic spin-Hall effect is not a general property,
since it is related to the specific form of the spin-orbit Hamiltonian.
It has been shown$^{20,22,23,24,25,26}$ that the static spin-Hall effect
exists for more complicated models involving higher-order (cubic) terms in
the momentum dependence of the spin-orbit Hamiltonian. The experimental
observation of the spin-Hall effect for the 2D hole system$^{27}$ described
by the spin-orbit Hamiltonian of this kind has confirmed this conclusion.
The edge spin accumulation due to the spin-Hall effect has been also
observed$^{8,11,28}$ for 2D electron systems. Recently, it has been
suggested$^{29,30}$ that deviation of electron band dispersion from the
parabolic one can lead to non-zero intrinsic static spin-Hall effect. This point,
however, remains controversial, since the analysis given in Ref. 17
predicts zero static spin-Hall effect even in this case. The calculations
presented in this paper also show the absence of static spin currents
for non-parabolic electron band dispersion.

In contrast to the static regime, the frequency-dependent induced spin
polarization and spin current have not been extensively studied.
Theoretical calculations of the frequency-dependent spin-Hall current based on
the Rashba model have been done in Refs. 16, 18 and 31 by using the methods
of non-equilibrium Green's functions, Kubo-Greenwood linear response theory,
and quantum kinetic equation, respectively, with the same result. The authors
of Ref. 31 also calculated the frequency dependence of the induced
spin polarization. A more complicated case of frequency-dependent response, when
both Rashba and Dresselhaus (linear in momentum) terms are included in the
spin-orbit Hamiltonian, has been studied in Refs. 32 and 33 in the dynamical
(collisionless) regime. A comparative numerical study of the frequency dependence
of the spin-Hall effect has been done in Ref. 23 for linear, cubic, and modified
Rashba models. The resonances in frequency dependence of spin-Hall conductivity
in magnetic field have been described in Ref. 34 for the Rashba model. Nevertheless,
a systematic investigation of the frequency-dependent problem of the induced
spin polarization and spin current is still missing.

A step towards systematic description of the frequency-dependent spin response
is undertaken in this paper by considering the important class of
spin-orbit Hamiltonians:
%1
\begin{equation}
\hat{h}_{\bf p}= \hbar \bOmega_{\bf p} \cdot \hat{\bsigma},~~~
\hbar \Omega^{\alpha}_{\bf p}= \xi_p \mu_{\alpha \beta} p_{\beta},
\end{equation}
where $\hat{\bsigma}$ is the vector of Pauli matrices, ${\bf p}=(p_x,p_y)$ is
the 2D momentum of electrons, and $\mu_{\alpha \beta}$ is the matrix of spin
velocities. Next, $\xi_p$ is an arbitrary function of the absolute value
of electron momentum. This function describes possible isotropic corrections to
spin-orbit interaction, which may have the same origin as the non-parabolicity
of the band spectrum. The case $\xi_p=1$ corresponds to the ${\bf p}$-linear
spin-orbit Hamiltonian of the general form. The ${\bf p}$-linear spin-orbit coupling
terms appear in quantum wells due to both the structural inversion asymmetry$^{14}$
(Rashba term) and bulk inversion asymmetry (Dresselhaus term), the latter
contribution is sensitive to orientation of the quantum well with respect to
crystallographic axes. By solving the quantum kinetic equation for the matrix
distribution function of electrons, with taking into account elastic
scattering, the spin polarization and spin currents are found on an equal footing
and a relation between them is established. The calculations not only provide
analytical expressions for these quantities, but also demonstrate a need to
reconsider the known results$^{5,35}$ for static spin polarization based on the
Hamiltonian (1) in special cases, when the symmetry of $\mu_{\alpha \beta}$
allows zero spin precession for some chosen directions of the spin vector. The
examples of this kind are the quantum wells grown in [001]
crystallographic direction in the case of equal Rashba and Dresselhaus spin
velocities and symmetric quantum wells (only the Dresselhaus term is
present) grown in [110] crystallographic direction. In particular, it is shown
that the induced spin polarization in these systems remains zero even when the
frequency of the applied field goes to zero. The calculations also show that the
static spin current in the electron systems described by the Hamiltonian (1) is
zero even if a non-parabolicity of electron band spectrum is taken into account.
The presented theory neglects the spin-orbit corrections to the
scattering potential, which means that the effects of extrinsic spin-orbit
coupling, such as the extrinsic spin currents and Elliot-Yafet spin relaxation,
are not included in the calculations.

The paper is organized as follows. In Sec. II we consider the quantum kinetic
equation and present its analytical solutions. The expressions for the induced spin
density vector and spin current tensor, obtained on the basis of these solutions,
are given in Sec. III. In that section we also establish a general relation between
these quantities and present a detailed analysis of two important cases, [001]-grown
and [110]-grown quantum wells with both Rashba and Dresselhaus spin-orbit coupling.
A relation between the induced spin density and electric current is derived and
analyzed in Sec. IV. The obtained results and the limits of their applicability
are discussed in Sec. V.

\section{General consideration}

The Hamiltonian of the problem is written in the form $\hat{H}_{p}+
\hat{h}_{\bf p}+ V_{\bf r} + \hat{H}^{ext}_{t}$, where $\hat{H}_{p}$ is
the Hamiltonian of free electrons in the absence of spin-orbit interaction,
$\hat{h}_{\bf p}$ is the spin-orbit Hamiltionian, $V_{\bf r}$ is the
potential of impurities or other static inhomogeneities (the spin-orbit
corrections to this potential are neglected, so only the intrinsic spin-orbit
coupling is considered), and $\hat{H}^{ext}_t= -e {\bf E}_t \cdot {\bf r}$ is
the Hamiltonian of the external perturbation due to the applied time-dependent
electric field ${\bf E}_t$ (here $e=-|e|$ is the electron charge).
The calculations are based on the quantum kinetic
equation for the Wigner distribution function $\hat{\rho}_{\mathbf{p}t}$, which
is a $2 \times 2$ matrix over the spin indices (see, for example, Refs. 36 and 37).
For the spatially-homogeneous problem considered below, this equation
is written in the form
%2
\begin{equation}
\frac{\partial\hat{\rho}_{\mathbf{p}t}}{\partial t}+\frac{i}{\hbar}\left[
\hat{h}_{\mathbf{p}},\hat{\rho}_{\mathbf{p}t}\right] + e{\bf E}_t \cdot
\frac{\partial\hat{\rho}_{\mathbf{p}t}}{\partial {\bf p}} =
\widehat{J} ( \hat{\rho}| \mathbf{p} t ),
\end{equation}
where $\widehat{J}$ is the collision integral describing the elastic scattering.
This integral is written below in the Markovian approximation and under
the assumptions $\hbar \omega \ll \overline{\varepsilon}$ and
$\hbar/\overline{\tau} \ll \overline{\varepsilon}$, where $\omega$ is the
frequency of the applied perturbation, $\overline{\varepsilon}$ is the
mean kinetic energy of electrons and $\overline{\tau}$ is the characteristic
scattering time. One has (see Ref. 37, problem 13.10)
%3
\begin{eqnarray}
\widehat{J}(\hat{\rho}|{\bf p} t)= \frac{1}{\hbar^2} \int
\frac{d {\bf p'}}{(2 \pi \hbar)^2} W(|{\bf p}-{\bf p}'|)
\int_{-\infty}^0 d t' e^{\lambda t'} ~~~~ \\
\times \! \left[
e^{i (\varepsilon_p+\hat{h}_{\bf p}) t'/\hbar} (\hat{\rho}_{{\bf p}' t}-\hat{\rho}_{{\bf p} t})
e^{-i(\varepsilon_{p'}+\hat{h}_{\bf p'} ) t'/\hbar} - ({\bf p}' \!\! \leftrightarrow \! {\bf p})
\right], \nonumber
\end{eqnarray}
where $W(|{\bf p}|)=\int d \Delta {\bf r} e^{-i {\bf p} \cdot \Delta {\bf r}/\hbar}
\left< \left< V_{{\bf r}+ \Delta {\bf r}} V_{\bf r}\right> \right>$
is the spatial Fourier transform of the correlation function
of the scattering potential, $\varepsilon_p$ is the kinetic energy of electron
in the absence of spin-orbit interaction (the energy spectrum is
isotropic but not necessarily parabolic), $\lambda \rightarrow +0$, and
$({\bf p}' \leftrightarrow {\bf p})$ denotes the term obtained from the
preceding one in the square brackets by permutation of momenta. The integration
over $t'$ in Eq. (3) is carried out elementary, but the resulting expression
is rather lengthy and, for this reason, is not presented here.

Searching for the linear response to the Fourier component ${\bf E} e^{-i \omega t}$
of the applied electric field, we represent the matrix distribution function
in the form $\hat{\rho}_{\mathbf{p}t}=\hat{f}^{(eq)}_{\mathbf{p}}+
\hat{f}_{\mathbf{p}} e^{-i\omega t}$, where $\hat{f}_{\mathbf{p}}$ is
the Fourier component of the non-equilibrium part of the distribution
function and $\hat{f}^{(eq)}_{\mathbf{p}}$ is the equilibrium distribution
function,
%4
\begin{eqnarray}
\hat{f}^{(eq)}_{\mathbf{p}} =
\frac{1}{2}[ f_{\varepsilon_p + \hbar \Omega_{\bf p} }+
f_{\varepsilon_p - \hbar \Omega_{\bf p} }] \nonumber \\
+ \frac{\bOmega_{\bf p} \cdot \hat{\bsigma}}{2 \Omega_{\bf p}}
[ f_{\varepsilon_p + \hbar \Omega_{\bf p} }-
f_{\varepsilon_p - \hbar \Omega_{\bf p} }],
\end{eqnarray}
which is expressed through the Fermi distribution $f_{\varepsilon}$.
Here and below, $\Omega_{\bf p} \equiv |{\bOmega_{\bf p}}|$.
The function (4) does not lead to spin polarization of electron
system because its matrix part is antisymmetric in momentum.
Since a substitution of the function (4) into Eq. (2) makes both
the commutator and the collision integral (3) equal to zero,
one has a closed integral equation for $\hat{f}_{\bf p}$:
%5
\begin{equation}
-i \omega \hat{f}_{\bf p} +\frac{i}{\hbar}\left[
\hat{h}_{\bf p}, \hat{f}_{\bf p} \right] + e{\bf E} \cdot
\frac{\partial \hat{f}^{(eq)}_{\bf p} }{\partial {\bf p}} =
\widehat{J} ( \hat{f}| \mathbf{p}),
\end{equation}
which determines the linear response of the electron system.

The induced spin density is defined as
%6
\begin{equation}
{\bf s}= \frac{1}{2} \int \frac{d {\bf p}}{(2 \pi \hbar)^2}
{\rm Tr}( \hat{\bsigma} \hat{f}_{\bf p}),
\end{equation}
and the induced (non-equilibrium) spin current density is given by the tensor
%7
\begin{equation}
q^{\alpha}_{\gamma}= \frac{1}{2} \int \frac{d {\bf p}}{(2 \pi \hbar)^2}
{\rm Tr}( \{ \hat{\sigma}_{\alpha}, \hat{u}_{\gamma}({\bf p}) \} \hat{f}_{\bf p} ),
\end{equation}
where $\hat{{\bf u}}({\bf p})=\partial (\varepsilon_p+\hat{h}_{\bf p})/\partial
{\bf p}$ is the group-velocity matrix, ${\rm Tr}$ denotes the matrix trace, and
$\{\hat{a},\hat{b}\}=(\hat{a} \hat{b}+\hat{b}\hat{a})/2$ denotes the symmetrized matrix
product. The expression (7) describes the flow of the spin polarized along $\alpha$
in the direction $\gamma$. One may also introduce the average spin ${\bf S}=
{\bf s}/n_{  2D}$, where $n_{  2D}=(2 \pi \hbar)^{-2} \int d {\bf p}
{\rm Tr} \hat{\rho}_{\bf p}$ is the electron density. The tensors
of spin polarizability, $\chi_{\alpha \beta}$, and spin conductivity,
$\Sigma^{\alpha}_{\gamma \beta}$, are introduced according to
%8
\begin{equation}
s_{\alpha} =\chi_{\alpha \beta}(\omega) E_{\beta},~~~
q^{\alpha}_{\gamma}=\Sigma^{\alpha}_{\gamma \beta}(\omega) E_{\beta}.
\end{equation}

It is assumed in the following that the spin-splitting energy
$2 \hbar \Omega_{\bf p}$ is small in comparison to the mean energy
of electrons. Then it is convenient to apply an efficient method of
solution of Eq. (5) based on the expansion of the collision integral
in series with respect to the small parameter $\hbar \Omega_{\bf p}
/\overline{\varepsilon}$; see Refs. 5, 25, 31, and problem 13.11 in the book 37.
Using the spin-vector representation $\hat{f}_{\bf p}={\rm f}^0_{\bf p}+
\hat{\bsigma} \cdot {\rm {\bf f}}_{\bf p}$ and retaining only the terms
of the first order in $\bOmega_{\bf p}$ under the collision integral,
we obtain coupled equations for scalar and vector parts of the
distribution function:
%9
\begin{eqnarray}
-i \omega {\rm f}^0_{\bf p} + \frac{1}{2}e{\bf E} \cdot
\frac{\partial [f_{\varepsilon_p + \hbar \Omega_{\bf p} }+
f_{\varepsilon_p - \hbar \Omega_{\bf p} }] }{\partial {\bf p}} \nonumber \\
= \frac{2 \pi}{\hbar} \int \! \! \frac{d {\bf p'}}{(2 \pi \hbar)^2}
W(|{\bf p}-{\bf p}'|) \left[ ({\rm f}^0_{\bf p'}-{\rm f}^0_{\bf p})
\delta (\varepsilon_{p'} -\varepsilon_{p}) \right. \nonumber \\
\left. - \hbar(\bOmega_{\bf p}-\bOmega_{\bf p'})
\cdot ({\rm {\bf f}}_{\bf p'}-{\rm {\bf f}}_{\bf p})
\frac{\partial \delta (\varepsilon_{p'} -\varepsilon_{p}) }{\partial
\varepsilon_{p'}} \right]
\end{eqnarray}
and
%10
\begin{eqnarray}
-i \omega {\rm {\bf f}}_{\bf p}
+ e {\bf E}  \cdot  \frac{\partial}{\partial {\bf p}} \frac{\bOmega_{\bf p}
[f_{\varepsilon_p + \hbar \Omega_{\bf p} }-
f_{\varepsilon_p - \hbar \Omega_{\bf p} }] }{2 \Omega_{\bf p} } \nonumber \\
- 2 [ \bOmega_{\bf p} \times {\rm {\bf f}}_{\bf p}]
= \frac{2 \pi}{\hbar} \int \! \! \frac{d {\bf p'}}{(2 \pi \hbar)^2}
W(|{\bf p}-{\bf p}'|) \left[ ({\rm {\bf f}}_{\bf p'}-{\rm {\bf f}}_{\bf p})
\right. \nonumber \\
\left. \times \delta (\varepsilon_{p'} -\varepsilon_{p})
- \hbar (\bOmega_{\bf p} \! -\bOmega_{\bf p'})
({\rm f}^0_{\bf p'} \! -{\rm f}^0_{\bf p}) \frac{\partial
\delta (\varepsilon_{p'} \! -\varepsilon_{p})}{\partial \varepsilon_{p'}} \right].
\end{eqnarray}
The expressions containing formal derivatives of the $\delta$-functions under
the integrals should be evaluated using integration by parts. The Fermi distribution
functions standing in the field terms also can be expanded in series of
$\Omega_{\bf p}$. Then one can see that the iterational expansions of
${\rm f}^0_{\bf p}$ and ${\rm {\bf f}}_{\bf p}$ start with the terms of
zero and first order in $\Omega_{\bf p}$, respectively. For this reason,
the last term under the collision integral in Eq. (9) can be neglected,
and the solution of this equation is
%11
\begin{equation}
{\rm f}^0_{\bf p} \simeq - \frac{e{\bf E} \cdot {\bf v}_{\bf p}}{
\nu^{  (1)}_{p}-i\omega} \frac{\partial
f_{\varepsilon_p}}{\partial \varepsilon_p},
\end{equation}
where ${\bf v}_{\bf p}= \partial \varepsilon_p/\partial {\bf p}$ is the
group velocity of electron in the absence of spin-orbit interaction. Here
and below, the relaxation rates appearing in the problem are defined as
%12
\begin{equation}
\nu^{  (n)}_{p} = \frac{2 \pi}{\hbar}
\int \! \! \frac{d {\bf p'}}{(2 \pi \hbar)^2} W(|{\bf p}-{\bf p}'|) [1-\cos (n \theta)]
\delta (\varepsilon_{p'} -\varepsilon_{p}),
\end{equation}
where $\theta$ denotes the scattering angle $\widehat{{\bf p}{\bf p}}'$.
The rate $\nu^{  (n)}_{p}$ describes relaxation of the $n$-th angular
harmonic of the distribution function.

Equation (11) describes the Drude response of the electron system. The
next correction to ${\rm f}^0_{\bf p}$ is of the order of
$(\hbar \Omega_{\bf p} /\overline{\varepsilon})^2$. This correction is
essential for calculation of the frequency-dependent conductivity and
dielectric function of 2D electrons with spin-orbit splitting$^{38,39}$
(see Sec. IV), but it is not important for calculation of the induced spin
polarization and spin current. After substituting the expression (11) into
the last term of the collision integral in Eq. (10), this term is unified
with the field term on the left-hand side of Eq. (10). As a
result, one gets a closed equation for the vector-function
${\rm {\bf f}}_{\bf p}$:
%13
\begin{eqnarray}
-i \omega {\rm {\bf f}}_{\bf p} - 2 [ \bOmega_{\bf p} \times
{\rm {\bf f}}_{\bf p}] + {\bf F}_{\bf p}(\omega) ~~~~~ \nonumber \\
= \frac{2 \pi}{\hbar} \int \! \! \frac{d {\bf p'}}{(2 \pi \hbar)^2}
W(|{\bf p}-{\bf p}'|) ( {\rm {\bf f}}_{\bf p'}
-{\rm {\bf f}}_{\bf p} ) \delta (\varepsilon_{p'} -\varepsilon_{p}).
\end{eqnarray}
The vector ${\bf F}$ contains both isotropic and anisotropic contributions:
%14-16
\begin{equation}
{\bf F}_{\bf p}(\omega)=e \bOmega_{\bf E} {\cal F}^{(i)}_{p}(\omega)-
e({\bf E} \cdot {\bf p}) \bOmega_{\bf p} {\cal F}^{(a)}_p(\omega),
\end{equation}
\begin{eqnarray}
{\cal F}^{(i)}_p(\omega)=R_p(\omega)
(\widetilde{\nu}_p-i\omega)- p^2 \frac{\partial
R_p(\omega)}{\partial p^2} \nu^{  (2)}_{p} , \\
{\cal F}^{(a)}_p(\omega)=\frac{2 \hbar}{p^2 \xi_p} \! \left[ R_p(\omega) \widetilde{\nu}_p
- p^2 \frac{\partial R_p(\omega)}{\partial p^2} (\nu^{  (2)}_{p} \! -i\omega) \right],
\end{eqnarray}
where $\bOmega_{\bf E}$ is the constant vector with components
$\Omega^{\alpha}_{\bf E}=\mu_{\alpha \beta} E_{\beta}$,
%17
\begin{equation}
R_p(\omega)= \frac{\xi_p}{\nu^{  (1)}_p-i\omega}
\frac{\partial f_{\varepsilon_p}}{\partial \varepsilon_p},
\end{equation}
and
%18
\begin{equation}
\widetilde{\nu}_p=\nu^{  (1)}_{p}-\nu^{  (2)}_{p} - \frac{p^2}{2}
\frac{\partial}{\partial \varepsilon_p} \left( \nu^{  (2)}_{p}
\frac{\partial \varepsilon_p}{\partial p^2} \right) \nonumber
\end{equation}
is a relaxation rate. Note that $\widetilde{\nu}_p$
goes to zero in the limit of short-range scattering potential.

It is convenient to search for the solution of Eq. (13) in the form
%19
\begin{equation}
{\rm {\bf f}}_{\bf p}=
\frac{e({\bf E} \cdot {\bf p}) \bOmega_{\bf p} {\cal F}^{(a)}_p(\omega)}{
\nu^{  (2)}_{p}-i\omega} + {\bf g}_{\bf p}.
\end{equation}
Since the first term of this expression is proportional to
$\bOmega_{\bf p}$, it does not contribute to the vector product in
Eq. (13), thereby representing a {\em non-precessing} part of the solution.
Note that the angular average of this term is directed along $\bOmega_{\bf E}$.
The substitution (19) leads to the following equation for ${\bf g}_{\bf p}$:
%20
\begin{eqnarray}
-i \omega {\bf g}_{\bf p} - 2 [ \bOmega_{\bf p} \times {\bf g}_{\bf p}] +
i \omega {\bf G}_p(\omega) \nonumber \\
= \frac{m_p}{\hbar^3}  \int_0^{2 \pi} \frac{d \varphi'}{2 \pi}
[W(|{\bf p}-{\bf p}'|) ({\bf g}_{\bf p'}-{\bf g}_{\bf p})]_{|{\bf p}'|=|{\bf p}|},
\end{eqnarray}
where $m_p=\frac{1}{2} (\partial p^2/\partial \varepsilon_p)$ is the
$p$-dependent effective mass and
%21
\begin{equation}
{\bf G}_p(\omega)=-\bOmega_{\bf E}
\frac{e \xi_p (\nu^{  (2)}_{p}+ \widetilde{\nu}_p-i\omega)}{(\nu^{  (1)}_{p}-i\omega)
(\nu^{  (2)}_{p}-i\omega)} \frac{\partial f_{\varepsilon_p}}{\partial \varepsilon_p}.
\end{equation}
The collision integral in Eq. (20) is already reduced, by means of
integration over the absolute value of ${\bf p}'$, to the integral
over the angle $\varphi'$ of the vector ${\bf p}'$. Owing to the
substitution (19), the inhomogeneous (field-dependent) term of Eq. (20),
$i \omega {\bf G}_p(\omega)$, is proportional to the
frequency $\omega$ and isotropic in the momentum space. With the aid
of the definition (21), it is convenient to write the angular-averaged
distribution function $\overline{ {\rm {\bf f}}}_{p}  \equiv (2 \pi)^{-1}
\int_0^{2 \pi} d \varphi {\rm {\bf f}}_{\bf p}$ as
%22
\begin{eqnarray}
\overline{{\rm {\bf f}}}_{p}=-{\bf G}_p(\omega)+ {\bf Q}_p(\omega)
+ \overline{{\bf g}}_{p}~, \\
{\bf Q}_p(\omega)=-e \bOmega_{\bf E} \frac{\partial[p^2 R_p(\omega)]}{\partial p^2}. \nonumber
\end{eqnarray}
We also point out the exact relation
%23
\begin{equation}
\overline{{\rm {\bf f}}}_{p}=-\frac{2}{i\omega} \overline{
[ \bOmega_{\bf p} \times {\bf g}_{\bf p} ] } + {\bf Q}_p(\omega),
\end{equation}
which is obtained by applying the procedure of angular averaging to Eq. (20)
and by using Eq. (22).

In spite of the isotropy of the term (21), Eq. (20) requires a numerical
solution. The physical reason for this is the effect of precession in the
presence of angular-dependent scattering. The angular dependence of the vector
product $[ \bOmega_{\bf p} \times {\bf g}_{\bf p}]$, in the general case, is
different from that of ${\bf g}_{\bf p}$ standing there, and the standard method
of solution, based on expansion of the distribution function in series of angular
harmonics, leads to an infinite set of coupled equations. There are, however,
a number of important situations when Eq. (20) can be solved analytically.
These situations are described in the subsections below.

\subsection{Short-range scattering potential}

Let us consider the limit of short-range scattering potential, when
$W(|{\bf p}-{\bf p}'|)$ is replaced by a constant $W$ and $\nu^{  (n)}_p=
\nu_p=m_pW/\hbar^3$ for any number $n$. The momentum dependence of the
scattering rate is associated with possible non-parabolicity of the
band spectrum and has to be ignored in the parabolic approximation.
Since the right-hand side of Eq. (20) is reduced in this case to
$\nu_p( \overline{{\bf g}}_{p}-{\bf g}_{\bf p} )$, a regular way
of solving exists. The solution is
%24
\begin{eqnarray}
{\bf g}_{\bf p}= -\frac{i \omega}{\nu_p (\nu_p-i\omega) \Delta^2_{\bf p} }
\frac{\partial f_{\varepsilon_p}}{ \partial \varepsilon_p} \left\{
(\nu_p-i\omega)^2 {\bf A}_{p} \right. \nonumber \\
\left. +  2 (\nu_p-i\omega)[\bOmega_{\bf p} \times {\bf A}_{p}] + 4 \bOmega_{\bf p}
(\bOmega_{\bf p} \cdot {\bf A}_{p}) \right\},
\end{eqnarray}
where $\Delta^2_{\bf p} =(\nu_p-i\omega)^2+4 \Omega^2_{\bf p}$ and
${\bf A}_{p}$ is the vector with components
%25
\begin{equation}
A^{\alpha}_{p}= e \xi_p (\hat{T}_p^{-1})_{\alpha \beta} \Omega^{\beta}_{\bf E}.
\end{equation}
Here $\hat{T}_p^{-1}$ denotes the matrix inverse of the symmetric $3 \times 3$ matrix
%26
\begin{equation}
T^{\alpha \beta}_p=4 \overline{\left[(\Omega^{\alpha}_{\bf p} \Omega^{\beta}_{\bf p}
-\delta_{\alpha \beta} \Omega^2_{\bf p})/\Delta^2_{\bf p} \right]}  +
\delta_{\alpha \beta} \frac{i\omega}{\nu_p}
\end{equation}
obtained as a result of angular averaging. The whole solution, according to
Eqs. (16)-(19), is
%27
\begin{equation}
{\rm {\bf f}}_{\bf p}=
-e({\bf E} \cdot {\bf p}) \bOmega_{\bf p} \frac{2 \hbar}{\xi_p} \frac{\partial}{\partial p^2}
\left[ \frac{\xi_p}{\nu_p-i\omega} \frac{\partial f_{\varepsilon_p}}{ \partial \varepsilon_p}
\right] + {\bf g}_{\bf p},
\end{equation}
and its angular average is written as
%28
\begin{equation}
\overline{{\rm {\bf f}}}_{p}= \frac{1}{\nu_p}\frac{\partial f_{\varepsilon_p}}{
\partial \varepsilon_p}\left(e \xi_p \bOmega_{\bf E} - \frac{i \omega}{\nu_p}
{\bf A}_{p} \right) + {\bf Q}_p(\omega).
\end{equation}
This vector determines the magnitude and the direction of the induced spin
polarization.

\subsection{Isotropic spin splitting}

The next exactly solvable situation is realized when the energy spectrum of
electrons remains isotropic in the presence of spin-orbit coupling. In other
words, $\Omega_{\bf p}=\Omega_{p}$ depends only on the absolute value of ${\bf p}$.
This imposes certain constraints on the matrix $\mu_{\alpha \beta}$:
%29
\begin{eqnarray}
\mu_{xx}^2+\mu_{yx}^2+\mu_{zx}^2 =\mu_{xy}^2+\mu_{yy}^2+\mu_{zy}^2, \nonumber \\
\mu_{xx} \mu_{xy}+\mu_{yx} \mu_{yy}+\mu_{zx} \mu_{zy}=0.
\end{eqnarray}
This situation is realized, for example, in [001]-grown quantum wells
($x~\| $[100], $y~\|$ [010], $z~\|$ [001]) with
only Rashba or only Dresselhaus type of spin-orbit coupling and in [111]-grown
quantum wells ($x~\|$ [11$\overline{2}$], $y~\|$ [$\overline{1}$10], $z~\|$ [111])
with both types of coupling. The solution is written in the form
similar to that of Eq. (24). By introducing the isotropic quantity
$\widetilde{\Delta}^2_{p} =(\nu^{  (1)}_p-i\omega)(\nu^{  (2)}_p-i\omega)+4
\Omega^2_{p}$ and the vector
%30
\begin{equation}
B^{\alpha}_{p}= e \xi_p (\hat{U}_p^{-1})_{\alpha \beta} \Omega^{\beta}_{\bf E},
\end{equation}
where $\hat{U}_p^{-1}$ denotes the matrix inverse of
%31
\begin{equation}
U^{\alpha \beta}_p=4 ( \overline{\Omega^{\alpha}_{\bf p} \Omega^{\beta}_{\bf p}}
-\delta_{\alpha \beta} \Omega^2_{p})/\widetilde \Delta^2_{p}  +
\delta_{\alpha \beta} \frac{i\omega}{\nu^{  (2)}_p},
\end{equation}
we find
%32
\begin{eqnarray}
{\bf g}_{\bf p}= -\frac{i \omega (\nu^{  (2)}_{p}+ \widetilde{\nu}_p-i\omega)}{\nu^{  (2)}_p
(\nu^{  (1)}_p-i\omega) (\nu^{  (2)}_p-i\omega) \widetilde{\Delta}^2_{p} }
\frac{\partial f_{\varepsilon_p}}{ \partial \varepsilon_p}
\nonumber \\
\times \left\{(\nu^{  (1)}_p-i\omega)(\nu^{  (2)}_p -i\omega){\bf B}_{p} \right.
\\ \left. +  2 (\nu^{  (2)}_p-i\omega)[\bOmega_{\bf p}
\times {\bf B}_{p}] + 4 \bOmega_{\bf p}
(\bOmega_{\bf p} \cdot {\bf B}_{p}) \right\}. \nonumber
\end{eqnarray}
It is easy to see that the results (24) and (32) become equivalent if one assumes
isotropic spin splitting in Eq. (24) and short-range scattering in Eq. (32).
The angular average of the whole solution is
%33
\begin{equation}
\overline{{\rm {\bf f}}}_{p}=\frac{(\nu^{  (2)}_{p}+ \widetilde{\nu}_p-i\omega)}{
\nu^{  (2)}_p (\nu^{  (1)}_{p}-i\omega)}\frac{\partial f_{\varepsilon_p}}{
\partial \varepsilon_p}\left(e \xi_p \bOmega_{\bf E} - \frac{i \omega}{\nu^{  (2)}_p}
{\bf B}_{p} \right)+ {\bf Q}_p(\omega).
\end{equation}

\subsection{Fixed precession axis}

There is also a special case, when Eq. (20) is solved in the most simple way.
This happens when the vector product $[ \bOmega_{\bf p} \times \bOmega_{\bf E}]$
is zero for arbitrary ${\bf p}$ and ${\bf E}$. In other words, the symmetry
of the matrix $\mu_{\alpha \beta}$ should allow existence of a fixed
(momentum-independent) precession axis. This imposes the following constraints:
%34
\begin{equation}
\mu_{xy}\mu_{yx}=\mu_{xx}\mu_{yy},~~~\mu_{xy}\mu_{zx}=\mu_{xx}\mu_{zy}.
\end{equation}
Under these conditions, both $\bOmega_{\bf p}$ and $\bOmega_{\bf E}$ are
directed along the fixed precession axis, without regard to directions of ${\bf p}$
and ${\bf E}$. The examples are [001]-grown quantum wells ($x~\|$ [100],
$y~\|$ [010], $z~\|$ [001]) with equal
absolute values of Rashba and Dresselhaus velocities, when the precession axis is in
the quantum well plane at the angle of $\pi/4$ or $-\pi/4$ with respect to the
main crystallographic axes, and [110]-grown wells ($x~\|$ [$\overline{1}$10],
$y~\|$ [001], $z~\|$ [110]) with only Dresselhaus type of coupling, when the
precession axis is perpendicular to the quantum well plane.
The solution of Eq. (20) in this case is non-precessing (makes the vector
product equal to zero) and isotropic (makes the collision integral equal to zero):
%35
\begin{equation}
{\bf g}_{\bf p}={\bf G}_p(\omega).
\end{equation}
This vector is directed along $\bOmega_{\bf E}$. The averaged whole
solution $\overline{{\rm {\bf f}}}_{p}$ is also directed along
$\bOmega_{\bf E}$:
%36
\begin{equation}
\overline{{\rm {\bf f}}}_{p}={\bf Q}_p(\omega).
\end{equation}

\subsection{Static limit}

If the frequency $\omega$ goes to zero, the solution of Eq. (20)
is trivial, ${\bf g}_{\bf p}=0$. Therefore, the function
${\rm {\bf f}}_{\bf p}$ from Eq. (19) with ${\bf g}_{\bf p}=0$
and $\omega=0$ describes the static spin-dependent response in the
general case. The only exception is the special case considered in
the previous subsection, when there exists a non-zero solution,
${\bf g}_{\bf p}={\bf G}_p(0)$. The averaged distribution function
for the static limit is
%37
\begin{equation}
\overline{{\rm {\bf f}}}_{p}= \bOmega_{\bf E}
\frac{e \xi_p (\nu^{  (2)}_{p}+ \widetilde{\nu}_p)}{
\nu^{  (1)}_p \nu^{  (2)}_{p}} \frac{\partial f_{\varepsilon_p}}{
\partial \varepsilon_p} + {\bf Q}_p(0)
\end{equation}
for the general case. In the special case only the last term of this
expression remains.

\section{Spin response}

To describe the spin response, one should calculate the integrals over momentum
${\bf p}$ in Eqs. (6) and (7). It is convenient to separate the angular averaging
from the integration over the squared absolute value of momentum, $p^2$, according
to $d {\bf p}=\frac{1}{2}~ \! d p^2 ~\! d \varphi$. Then, after using the representation
$\hat{f}_{\bf p}={\rm f}^0_{\bf p}+ \hat{\bsigma} \cdot {\rm {\bf f}}_{\bf p}$
and taking the matrix trace, the density of the induced spin polarization
is given by
%38
\begin{equation}
{\bf s}= \int_0^{\infty} \frac{d p^2}{4 \pi \hbar^2} \overline{{\rm {\bf f}}}_{p}~,
\end{equation}
and the density of non-equilibrium spin current is
%39
\begin{eqnarray}
q^{\alpha}_{\gamma}= \int_0^{\infty} \frac{d p^2}{4 \pi \hbar^2} [ ~ \!
m_p^{-1} \overline{{\rm f}^{\alpha}_{\bf p} p_{\gamma}}  \nonumber \\
 + \mu_{\alpha \beta} \overline{{\rm f}^0_{\bf p} \left( \delta_{\beta \gamma}
\xi_p  + 2(\partial \xi_p/\partial p^2) p_{\beta} p_{\gamma} \right)} ~ \! ].
\end{eqnarray}
The second term, which appears in the expression (39) owing to the spin-orbit
correction to the group velocity, gives zero contribution because, according
to Eq. (11), the scalar part of the distribution function is antisymmetric
in momentum. If the frequency $\omega$ is zero, the vector part of the distribution
function is symmetric in momentum, so the first term of the expression (39) also
gives zero contribution. Therefore, the non-equilibrium static spin currents
do not exist for the model described by the spin-orbit Hamiltonian (1). At non-zero
frequency, the spin currents are associated with ${\bf g}$-contribution to the
distribution function, because the first term of Eq. (19) is symmetric in ${\bf p}$:
%40
\begin{equation}
{\bf q}_{\gamma}= \int_0^{\infty} \frac{d p^2}{4 \pi \hbar^2 m_p}
\overline{{\bf g}_{\bf p} p_{\gamma}},
\end{equation}
where ${\bf q}_{\gamma}=(q_{\gamma}^x,q_{\gamma}^y,q_{\gamma}^z)$.
Looking at the expressions (24) and (32), one can
conclude that it is the second terms in the braces of these expressions
that are responsible for the spin currents.

On the other hand, the induced spin density is determined by the angular-averaged
symmetric part of ${\rm {\bf f}}_{\bf p}$ and exists in the static regime as well.
Applying either the exact relations (22) and (23) or the expressions (28),(33),
(36), and (37) describing different physical situations, one should always ignore
the term ${\bf Q}_p(\omega)$ which represents a full derivative over $p^2$ and, for
this reason, does not contribute to the integral (38). The static spin polarization
obtained in this way is
%41
\begin{eqnarray}
{\bf s}= \frac{e \bOmega_{\bf E} }{4 \pi \hbar} \int_0^{\infty} \! d p^2 \xi_p
\frac{\partial f_{\varepsilon_p}}{\partial \varepsilon_p}
\left[ \frac{1}{\nu^{  (2)}_{p}} \right. \nonumber \\
\left. - \frac{p^2}{2 \nu^{  (1)}_{p} \nu^{  (2)}_{p}}
\frac{\partial}{\partial \varepsilon_p} \left( \nu^{  (2)}_{p}
\frac{\partial \varepsilon_p}{\partial p^2} \right) \right].
\end{eqnarray}
In the parabolic approximation $\varepsilon_p=p^2/2m$ and at $\xi_p=1$ this
expression is rewritten as
%42
\begin{equation}
{\bf s}= \bOmega_{\bf E} \frac{e m}{2 \pi \hbar}
\int_0^{\infty} d \varepsilon \frac{\partial f_{\varepsilon}}{\partial
\varepsilon} \left[ \tau^{  (2)}_{\varepsilon} +
\tau^{  (1)}_{\varepsilon} \frac{1}{2} \frac{\partial \ln \tau^{  (2)}_{\varepsilon}
}{\partial \ln \varepsilon} \right],
\end{equation}
where $\tau^{  (n)}_{\varepsilon_p}=1/\nu^{  (n)}_{p}$ are the relaxation times.
For degenerate electron gas, when $(\partial f_{\varepsilon}/\partial \varepsilon) \simeq
-\delta(\varepsilon-\varepsilon_{  F})$ and $\varepsilon_{  F}$ is the Fermi
energy, the integral over energy is taken in a straightforward way.
In the case of short-range scattering potential
($\tau^{  (n)}_{\varepsilon}=\tau=1/\nu$) one arrives at the well-known result$^{35}$
$s_{\alpha} =\chi_{\alpha \beta} E_{\beta}$ with $\chi_{\alpha \beta}=-\mu_{\alpha
\beta} [e m \tau/2 \pi \hbar^2]$ valid for any linear spin-orbit Hamiltonian
including the case of Rashba spin-orbit coupling studied in the early papers.$^{2,4}$

It is important that Eqs. (41) and (42) are not valid in the special
situation when a fixed precession axis exists; see subsection C of Sec. II.
In fact, a straightforward substitution of Eq. (36) into Eq. (38) shows
that, at arbitrary frequency of the applied field, the spin polarization does not
appear in this special situation. The physical explanation of this remarkable
property is based on the facts that in the case of a fixed precession axis (a)
there is no spin relaxation$^{40}$ by the Dyakonov-Perel-Kachorovskii (DPK)
mechanism$^{41}$ for the spins directed along this axis and (b) the induced
spin polarization can, in principle, appear only along this axis, without
regard to the direction of the applied electric field. Imagine that the
electric field is abruptly turned on.
The anisotropic distribution of electrons over momenta, which determines the
electric current, is established during the momentum relaxation time
$1/\nu_p^{  (1)}$. However, the distribution of electrons over spins, which
determines the spin density, is established during the spin relaxation time,
and the corresponding transient process becomes infinitely slow if the spin
relaxation is absent. Therefore, if a periodic alternating field acts on
electrons in a sample with a fixed precession axis, the spin density cannot
react to the field at any frequency $\omega$, and the spin polarizability
is zero. The absence of the static spin polarization in the case of a
fixed precession axis cannot be revealed by consideration
of the static spin response alone. From the formal point of view, this paradox
is related to the fact that in the static limit the additional part of the
distribution function, ${\bf g}_{\bf p}={\bf G}_p(0)$, which cancels the
contribution of the first term in the expression (19), still exists,
{\em only} in the case of a fixed precession axis. One can say that the
dependence of the induced spin polarization on the parameters (components
$\mu_{\alpha \beta}$) of the spin-orbit Hamiltonian is non-analytic
in the region of parameters where the fixed precession axis appears.
This means that the result for the spin polarization depends on the order
of limiting transitions. If first $\omega$ is aimed to zero and then
the precession axis is fixed, the polarization is finite. If first
the precession axis is fixed and then $\omega$ is aimed to zero, the
polarization is zero. More details on the issue of non-analyticity will be given
below in this section, by considering concrete examples. It is important to state
that the consideration given above does not provide spin relaxation mechanisms
other than the DPK mechanism. Inclusion of the Elliot-Yafet mechanism
(see Refs. 42 and 43 for the 2D case) can lead to a finite relaxation of the
spins oriented along the fixed precession axis and, therefore, to a finite
induced spin polarization for this special case; see Sec. V for more
discussion.

Using Eq. (23) together with Eqs. (20), (38), and (40), one can obtain an exact
relation connecting the induced spin polarization and spin current. From
Eqs. (38) and (23) one has ${\bf s}=-(2 \pi i \omega)^{-1} \int_0^{\infty} d p^2
\overline{[ \bOmega_{\bf p} \times {\bf g}_{\bf p} ] }$.
Since $[ \bOmega_{\bf p} \times {\bf g}_{\bf p} ]=\xi_p [\bmu_{\beta}
\times {\bf g}_{\bf p}] p_{\beta}$, where $\bmu_{\beta}$ is the vector
with components $\mu_{\alpha \beta}$, we find, for the case of parabolic
band and $\xi_p=1$,
%43
\begin{equation}
{\bf s}= -\frac{2 m}{i \hbar \omega} [\bmu_{\beta} \times {\bf q}_{\beta}].
\end{equation}
This relation is remarkable in the sense that it does not contain any parameters
describing scattering. The relation (43) is also valid for a non-parabolic model
provided that the momentum dependence of $1/m_p$ and $\xi_p$ is the same; in
this case the effective mass $m$ should be replaced by the momentum-independent
quantity $m_p \xi_p$. Using Eq. (43), one can directly relate the {\em low-frequency}
behavior of the spin current to the {\em static} induced spin polarization.
Equation (43) can be derived from the continuity equation (for the
case of Rashba model see Ref. 44 and Sec. 64 of Ref. 37). In detail,
when Eq. (2) is summed over the momentum ${\bf p}$, the
collision-integral contribution and the electric-field term vanish, and Eq. (43)
directly follows from the relation between the momentum-averaged first and
second terms on the left-hand side of Eq. (2).
%Though this method requires more general definitions of the
%spin polarization and spin current densities, namely Eqs. (6) and (7) with
%$\hat{f}_{\bf p}$ replaced by $\hat{\rho}_{\bf p}$, the contribution
%of the equilibrium spin currents$^{12}$ remains non-essential, since it
%vanishes in the vector product in the right-hand side of Eq. (43).
This observation also demonstrates that the validity of Eq. (43) is not restricted
by the assumption of linear response used in this paper: the non-equilibrium part of
the distribution function $\hat{\rho}$ is not necessarily small. On the other hand,
Eq. (43) is specific for the chosen form (1) of the spin-orbit Hamiltonian.
Applying Eq. (43) to the model including both Rashba and Dresselhaus linear spin-orbit
coupling terms in [001]-grown quantum wells (see subsection A below), one obtains
the relations between the components of the spin density and spin current recently
derived$^{18,32}$ for this model. The corresponding relations$^{37,44}$ for the
Rashba model also follow from Eq. (43).

Since it is established that the non-parabolic corrections to the band
spectrum and the corrections to the spin-orbit Hamiltonian do not lead to
qualitative modifications of the induced spin polarization and spin current, we
neglect these corrections in the following, by assuming $\varepsilon_p=p^2/2m$
and $\xi_p=1$. Let us consider first the frequency behavior of the induced spin
polarization and spin current in the quantum wells described by the Rashba model.
The non-zero components of the spin-velocity matrix are $\mu_{xy}=-\mu_{yx}=
v_{  R}$, where $v_{  R}$ is the Rashba velocity. The spin splitting described
by the Hamiltonian (1) is isotropic in this situation. Using Eq. (33) and taking
into account that the matrix (31) is diagonal, we obtain
%44
\begin{equation}
{\bf s}= - \bOmega_{\bf E} \frac{em}{2 \pi \hbar} {\cal T}(\omega,v_{  R}),
\end{equation}
where the frequency-dependent function
%45
\begin{eqnarray}
{\cal T}(\omega,v)= \int_0^{\infty} d \varepsilon_p
\left(-\frac{\partial f_{\varepsilon_p}}{\partial \varepsilon_p} \right)
\left(1-\frac{\varepsilon_p}{2} \frac{\partial \nu^{  (2)}_p/\partial
\varepsilon_p}{\nu^{  (1)}_p-i\omega} \right) \nonumber \\
\times \frac{2 (vp/\hbar)^2}{2 (vp/\hbar)^2 \nu^{  (2)}_p
\!\! -i\omega[(\nu^{  (1)}_p \!\! -i\omega)(\nu^{  (2)}_p \!\! -i\omega)+4(vp/\hbar)^2] }
\end{eqnarray}
has dimensionality of time. Equations (44) and (45) generalize the result
of Ref. 31 obtained in the limit of short-range scattering potential to
the case of arbitrary scattering potential. They also describe [111]-grown
quantum wells with both Rashba and Dresselhaus spin-orbit coupling, where
$\mu_{xy}=-\mu_{yx}=v_{  R}+ 2 v_{  D}/\sqrt{3}$ and $v_{  D}$
is the Dresselhaus velocity. In this case, $v_{  R}$ in Eq. (44) should
be merely replaced by $v_{  R}+ 2 v_{  D}/\sqrt{3}$ and the direction
of the spin polarization vector ${\bf s}$ with respect to ${\bf E}$ is
the same as for the pure Rashba coupling. In particular, ${\bf s}=[em v
{\cal T}(\omega,v)/2 \pi \hbar^2](-E_y,E_x,0)$ where $v$ is either
$v_{  R}$ or $v_{  R}+ 2 v_{  D}/\sqrt{3}$, respectively.
Next, in symmetric [001]-grown quantum wells, where only the Dresselhaus
spin-orbit coupling is present and the non-zero components are $\mu_{yy}=-
\mu_{xx}=v_{  D}$, Eq. (44) with ${\cal T}(\omega,v_{  D})$ is also
valid, leading to ${\bf s}=[em v_{  D} {\cal T}(\omega,v_{  D})/2 \pi
\hbar^2](E_x,-E_y,0)$. In all these cases, the spin polarization is in the
quantum well plane ($s_z=0$), and the direction of this polarization is
frequency-independent. The spin currents can be either calculated directly
or extracted from Eq. (43). There exist only the currents of $z$-polarized
spins, and these currents are directed perpendicular to the applied field:
%46
\begin{equation}
\left( \begin{array}{c} q^z_x \\ q^z_y \end{array} \right)
= \frac{i e \omega {\cal T}(\omega,v)}{4 \pi \hbar} \left( \begin{array}{c}
-E_y \\ E_x \end{array} \right),
\end{equation}
where ${\cal T}(\omega,v)$ is given by Eq. (45) with $v=v_{  R}$
for the pure Rashba coupling and $v=v_{  R}+ 2 v_{  D}/\sqrt{3}$
for [111]-grown quantum wells. For symmetric [001]-grown quantum wells
one should use $v=v_{  D}$ and change the sign of the right-hand side.
Therefore, the symmetry properties and frequency dependence of the spin currents
remain the same for all important cases of isotropic spin splitting considered in
this paragraph. In the limit of short-range scattering potential and in the case of
degenerate electrons, Eqs. (46) and (45) with $v=v_{  R}$ give the result
obtained previously in Ref. 16. The universal behavior$^{13}$ of the spin-Hall
conductivity $\Sigma^{z}_{xy}=-\Sigma^z_{yx}$ exists in the limit
$v p_{  F}/\hbar \gg \omega \gg \nu^{  (2)}_{p_{  F}}$, where
$p_{  F}$ is the Fermi momentum.

Now we turn to more complicated situations when the anisotropy of spin
splitting is essential due to combined effect of both Rashba and
Dresselhaus spin-orbit coupling. These cases are considered in the
following subsections in the limit of short-range scattering potential,
when the expressions obtained in subsection A of Sec. II are valid.

\subsection{[001]-grown quantum wells}

Let us study the case of [001]-grown quantum wells. If the Cartesian
coordinate axes are chosen along the principal crystallographic
directions, there are four components of the spin-velocity tensor:
%47
\begin{equation}
\mu_{xy}=-\mu_{yx}=v_{  R},~~\mu_{yy}=-\mu_{xx}=v_{  D}.
\end{equation}
The spin splitting depends on the angle $\varphi$ of the vector ${\bf p}$ according to
$\Omega_{\bf p}^2=(v_{  R}^2+v_{  D}^2)(p/\hbar)^2-2v_{  R} v_{  D} (p/\hbar)^2
\sin(2 \varphi)$ and $\Delta^2_{\bf p}=a-b \sin(2 \varphi)$, where
%48
\begin{equation}
a=(\nu-i\omega)^2+4(v_{  R}^2+v_{  D}^2)(p/\hbar)^2,~~b=8 v_{  R}
v_{  D} (p/\hbar)^2.
\end{equation}
The angular averaging in Eq. (26) results in $T_p^{xx}=T_p^{yy}=
(\nu-i\omega)^2/2r - 1/2 + i\omega/\nu$ and
$T_p^{yx}=T_p^{xy}=b/2r -(a/r - 1)(v_{  R}^2+v_{  D}^2)/4
v_{  R}v_{  D}$, where $r=\sqrt{a^2-b^2}$.
After some transformations, we obtain the relation defining the tensors
of spin polarizability and spin conductivity:
%49
\begin{eqnarray}
\left( \begin{array}{c} s_x \\ s_y \end{array} \right) =
\frac{em}{2 \pi \hbar^2 \nu} \int_0^{\infty} \frac{d \varepsilon_p}{D_p(\omega)}
\left(-\frac{\partial f_{\varepsilon_p}}{\partial \varepsilon_p} \right)
\nonumber \\
\times \left[ \begin{array}{cc} \kappa_p(\omega) & -\widetilde{\kappa}_p(\omega) \\
\widetilde{\kappa}_p(\omega) & -\kappa_p(\omega)  \end{array} \right]
\left( \begin{array}{c} E_x \\ E_y \end{array} \right),
\end{eqnarray}
and
%50
\begin{eqnarray}
\left( \begin{array}{c} q^z_x \\ q^z_y \end{array} \right) =
\frac{i e \omega}{4 \pi \hbar \nu} \frac{v_{  R}^2-v_{  D}^2}{4 v_{  R} v_{  D}}
\int_0^{\infty} \frac{d \varepsilon_p}{D_p(\omega)}
\left(-\frac{\partial f_{\varepsilon_p}}{\partial \varepsilon_p} \right)
\nonumber \\
\times \left[ \begin{array}{cc} \theta_p(\omega) & -\widetilde{\theta}_p(\omega) \\
\widetilde{\theta}_p(\omega) & -\theta_p(\omega)  \end{array} \right]
\left( \begin{array}{c} E_x \\ E_y \end{array} \right),
\end{eqnarray}
where the denominator is given by
%51
\begin{eqnarray}
D_p(\omega)=\frac{(v_{  R}^2-v_{  D}^2)^2}{8 v^2_{  R} v^2_{  D}}
\left( \frac{a}{r} - 1 \right) \nonumber \\
+ \frac{i \omega}{\nu} \left( \frac{(\nu-i\omega)^2}{r} - 1 \right)
+ \left( \frac{i \omega}{\nu} \right)^2.
\end{eqnarray}
The elements of the matrix in Eq. (49) are
%52
\begin{eqnarray}
\kappa_p(\omega)=\frac{v_{  R}^2-v_{  D}^2}{4 v_{  D}}
\left\{ \left[\frac{v_{  R}^2-v_{  D}^2}{2v_{  R}^2} -
\frac{i \omega}{\nu} \right] \right. \nonumber \\
\left. \times \left(\frac{a}{r} - 1 \right) + \frac{i \omega}{\nu}
\frac{v_{  D}}{v_{  R}} \frac{b}{r} \right\},
\end{eqnarray}
and $\widetilde{\kappa}_p(\omega)$ is obtained from this expression
by the permutations $v_{  R} \leftrightarrow v_{  D}$.
The elements of the matrix in Eq. (50) are given by
%53
\begin{equation}
\theta_p(\omega)=
\left(1- \frac{i \omega}{\nu} \right) \left(\frac{a}{r} - 1 \right)
\end{equation}
and
%54
\begin{equation}
\widetilde{\theta}_p(\omega)=
\frac{v_{  R}^2+v_{  D}^2}{2 v_{  R} v_{  D}}
\left(\frac{a}{r} - 1 \right) - \frac{i \omega}{\nu}
\frac{b}{r}.
\end{equation}
Equations (49) and (50) demonstrate the symmetry relations
$\chi_{xx}=-\chi_{yy}$, $\chi_{xy}=-\chi_{yx}$, $\Sigma^z_{xx}
=-\Sigma^z_{yy}$, and $\Sigma^z_{xy}=-\Sigma^z_{yx}$.

Though the matrix in Eq. (49) retains the symmetry of $\mu_{\alpha \beta}$,
the ratio $\widetilde{\kappa}_p(\omega)/\kappa_p(\omega)$ is not equal
to $v_{  R}/v_{  D}$. For this reason, the direction of spin
polarization in the plane is different from the direction of
$\bOmega_{\bf E}$ and depends on the frequency. The direction
of the spin current is not perpendicular to the direction
of the field and is also frequency-dependent. The components
of the spin density and spin current are related according to
Eq. (43), which can be written, for this particular case, in
the form$^{18,32}$ $q^z_x=(i \hbar \omega/2m)(v_{  R} s_x+v_{  D}
s_y)/(v_{  R}^2-v_{  D}^2)$ and $q^z_y=(i \hbar \omega/2m)(v_{  D} s_x
+ v_{  R} s_y)/(v_{  R}^2-v_{  D}^2)$.

In the collisionless limit, $\nu \rightarrow 0$, Eqs. (49) and (50)
are reduced to the results obtained in Ref. 32. Note that the formal
substitution $\nu \rightarrow 0$ makes the spin currents finite at $\omega
\rightarrow 0$ because the denominator $D_p(\omega)$ is reduced to
$-(\omega/\nu)^2$, while the functions (53) and (54) become proportional
to $\omega/\nu$. The static spin currents for the systems described
by the spin-velocity matrix (47) have been also studied in the
collisionless limit in Refs. 45 and 46. All these studies show that,
as the ratio $v_{  R}/v_{  D}$ is varied, the spin current reverses
its sign going through zero at $v_{  R}^2=v_{  D}^2$. This general
property, also reflected in Eq. (50), follows from the Berry phase
analysis.$^{45}$

\begin{figure}[ht]
\begin{center}
\includegraphics[scale=0.4]{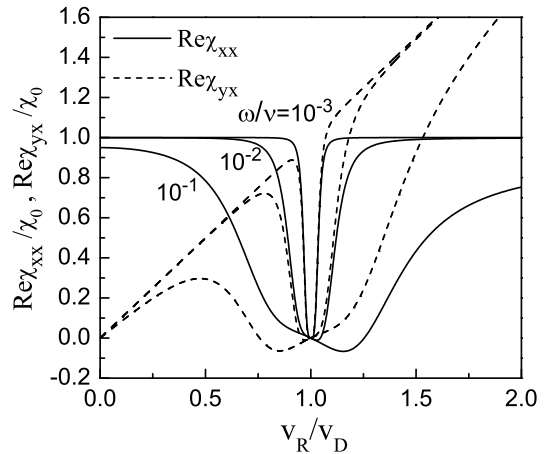}
\end{center}
\addvspace{-0.7 cm}\caption{Dependence of the real part of
spin polarizability in [001]-grown quantum wells on the ratio
$v_{  R}/v_{  D}$ at $v_{  D}p_{  F}/\hbar \nu=1$ for
several frequencies. The polarizability is expressed
in the units of $\chi_0=em v_{  D}/2 \pi \hbar^2 \nu$. The diagonal
and off-diagonal components are shown by the solid and dashed
lines, respectively.}
\end{figure}

\begin{figure}[ht]
\begin{center}
\includegraphics[scale=0.4]{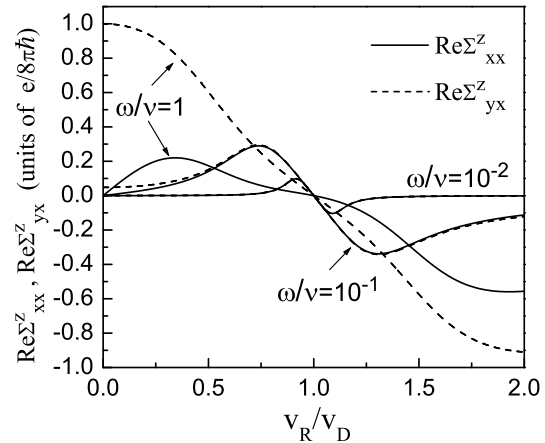}
\end{center}
\addvspace{-0.7 cm}\caption{Dependence of the real part of
spin conductivity in [001]-grown quantum wells on the ratio $v_{  R}/v_{  D}$ at
$v_{  D}p_{  F}/\hbar \nu=1$ for several frequencies.
The spin conductivity is expressed in the universal
units. The diagonal and off-diagonal (Hall) components are
shown by the solid and dashed lines, respectively. The
difference between them is negligible at small $\omega$. }
\end{figure}

In the case $v_{  R}^2=v_{  D}^2$, as already mentioned,
there exists a fixed precession axis directed at the angle of $\pi/4$
(or $-\pi/4$) in the quantum well plane. Therefore, not only the spin
current, but also the induced spin density goes to zero in this case.
This behavior is demonstrated in Figs. 1-3, where the calculated real parts
of the components of spin polarizability and spin conductivity tensors,
${\rm Re} \chi_{\alpha \beta}$ and ${\rm Re} \Sigma^z_{\alpha \beta}$,
are plotted as functions of the ratio of Rashba and Dresselhaus
velocities. The case of degenerate electron gas is assumed.
The spin polarizability is expressed in the units
of static polarizability for symmetric [001]-grown quantum wells,
$\chi_0=em v_{  D}/2 \pi \hbar^2 \nu$. If the frequency decreases,
both $\chi_{xx}$ and $\chi_{yx}$, as expected, approach their
static values $\chi_0$ and $(v_{  R}/v_{  D})\chi_0$, respectively.
However, this never happens at $v_{  R} = v_{  D}$, when the
polarizability remains exactly zero at arbitrary $\omega$. The real
part of the spin conductivity (Fig. 2) shows prominent peaks near the point
$v_{  R} = v_{  D}$ at small $\omega$, though its behavior is analytic.
The imaginary parts of $\chi_{xx}$ and $\chi_{yx}$ (not shown) also have
sharp peaks in the vicinity of $v_{  R} = v_{  D}$ at small $\omega$.
The depression of the spin polarizability in the region $v_{  R} \simeq
v_{  D}$ is extended with the increase of the disorder, as shown in Fig. 3.
This also means that, especially for the "dirty" case $v_{  D}
p_{  F} \ll \hbar \nu$, the frequency dispersion of the spin
polarizability remains significant at very small frequencies if
$v_{  R}$ is close enough to $v_{  D}$. The corresponding behavior
is illustrated in Fig. 4 and is explained by the reduction and
disappearance of the DPK spin relaxation as $v_{  R}$ approaches
to $v_{  D}$.

\begin{figure}[ht]
\begin{center}
\includegraphics[scale=0.4]{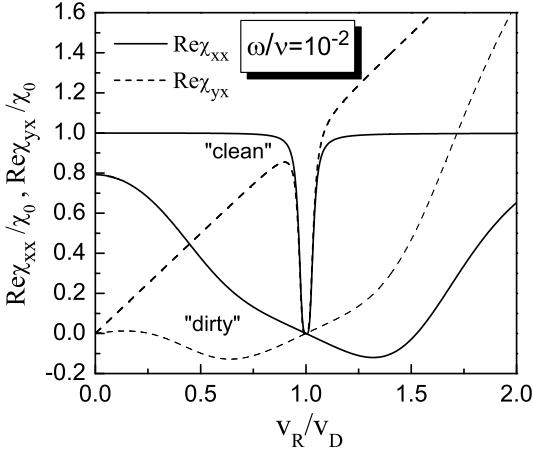}
\end{center}
\addvspace{-0.7 cm}\caption{Effect of disorder on the
real part of spin polarizability. The "clean" and the "dirty"
cases correspond to $v_{  D}p_{  F}/\hbar \nu=10$ and
$v_{  D}p_{  F}/\hbar \nu=0.1$, respectively.}
\end{figure}

\begin{figure}[ht]
\begin{center}
\includegraphics[scale=0.4]{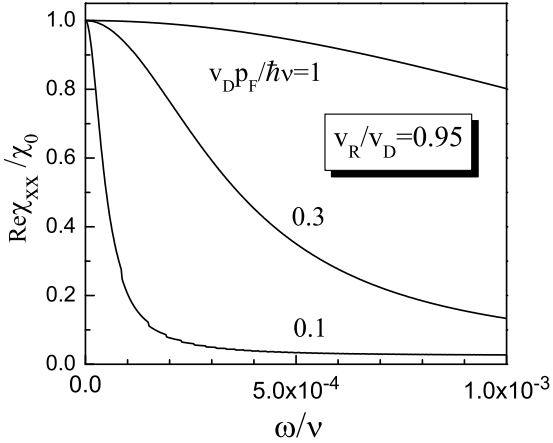}
\end{center}
\addvspace{-0.7 cm}\caption{Low-frequency dispersion of the
real part of spin polarizability in [001]-grown quantum wells
in the vicinity of the fixed precession axis ($v_{  R}$ is
close to $v_{  D}$).}
\end{figure}

\subsection{[110]-grown quantum wells}

The next case we consider is [110]-grown quantum wells. If the Cartesian
coordinate axes are chosen with $OZ$ perpendicular to the quantum well plane and
$OY$ along the principal crystallographic direction ($x~\|$ [$\overline{1}$10],
$y~\|$ [001], $z~\|$ [110]), there are three
components of the spin-velocity tensor:
%55
\begin{equation}
\mu_{xy}=-\mu_{yx}=v_{  R},~~\mu_{zx}=v_{  D}/2.
\end{equation}
One has $\Omega_{\bf p}^2=(v_{  R}^2+v_{  D}^2/8)(p/\hbar)^2+
(v_{  D}^2/8) (p/\hbar)^2
\cos(2 \varphi)$ and $\Delta^2_{\bf p}=c+d \cos(2 \varphi)$, where
%56
\begin{equation}
c=(\nu-i\omega)^2+(4 v_{  R}^2 + v_{  D}^2/2)(p/\hbar)^2,~~d=v_{  D}(p/\hbar)^2/2.
\end{equation}
Applying the equations listed in subsection A of Sec. II, we find that
the field $E_x$ can induce both $y$- and $z$-polarized spins, while the
field $E_y$ induces only $x$-polarized spins:
%57
\begin{equation}
s_x = - \frac{em v_{  R} E_y}{2 \pi \hbar^2 \nu} \int_0^{\infty} \! \! d \varepsilon_p
\left(-\frac{\partial f_{\varepsilon_p}}{\partial \varepsilon_p} \right)
\left(1- \frac{i \omega/\nu}{K^{x}_p(\omega)} \right),
\end{equation}
%58
\begin{equation}
s_y = \frac{em v_{  R} E_x}{2 \pi \hbar^2 \nu} \int_0^{\infty} \! \! d \varepsilon_p
\left(-\frac{\partial f_{\varepsilon_p}}{\partial \varepsilon_p} \right)
\left(1- \frac{i \omega/\nu}{K^{y}_p(\omega)} \right),
\end{equation}
and $s_z=-(v_{  D}/2 v_{  R})s_y$, where
%59
\begin{equation}
K^{x}_p(\omega)= \left( \frac{4 v_{  R}^2}{v_{  D}^2} + 1 \right)
\left(\sqrt{\frac{c-d}{c+d}}-1\right) + \frac{i \omega}{\nu} ,
\end{equation}
and
%60
\begin{equation}
K^{y}_p(\omega)= - \frac{4 v_{  R}^2}{v_{  D}^2}
\left(\sqrt{\frac{c+d}{c-d}}-1 \right) + \frac{i \omega}{\nu}.
\end{equation}
The component $s_z$ is a symmetric function of $v_{  R}$, while
$s_x$ and $s_y$ are antisymmetric functions of $v_{  R}$.

The spin currents appear for $y$- and $z$-polarized spins and flow
in the direction perpendicular to the applied field:
%61
\begin{equation}
q^y_x = \frac{i \hbar \omega}{m v_{  D} (4 v_{  R}^2/v_{  D}^2 +1)}
s_x,~~q^y_y = \frac{i \hbar \omega v_{  D}}{4 m v_{  R}^2} s_y,
\end{equation}
and
%62
\begin{equation}
q^z_x = (2v_{  R}/v_{  D}) q^y_x ,~~~ q^z_y=(2v_{  R}/v_{  D})q^y_y.
\end{equation}
The relations (61) and (62) satisfy the general requirement (43).
The components $q^z_x$ and $q^z_y$ are symmetric functions of $v_{  R}$,
while $q^y_x$ and $q^y_y$ are antisymmetric functions of $v_{  R}$.

For symmetric quantum wells, where $v_{  R}=0$, both the polarization
and the spin currents disappear for arbitrary $\omega$. If both
$v_{  R}$ and $\omega$ go to zero, the $z$-component of the spin
polarizability tensor, $\chi_{zx}(\omega)$, is described by a
simple formula applied to the case of degenerate electrons:
%63
\begin{equation}
\chi_{zx} \simeq -\frac{\chi_0}{2} \frac{(2v_{  R}/v_{  D})^2 \gamma_{  F}}{
(2v_{  R}/v_{  D})^2 \gamma_{  F} -i\omega},
\end{equation}
where $\gamma_{  F}=\sqrt{\nu^2+ (v_{  D}p_{  F}/\hbar)^2}-\nu$.
Equation (63) illustrates the non-analytic behavior of the spin polarizability
in the vicinity of the fixed precession axis. The calculated dependence of
the real part of $\chi_{zx}$ on the ratio $v_{  R}/v_{  D}$, shown in
Fig. 5 for the case of degenerate electron gas, is similar to the dependence
of $\chi_{xx}$ shown in Fig. 1 for [001]-grown wells. The region of depression
near $v_{  R}=0$ is extended in the "dirty" limit $v_{  D}p_{  F}
\ll \hbar \nu$, as it is seen directly from Eq. (63). The behavior of the
component $\chi_{yx}$ remains analytic at $v_{  R} \rightarrow 0$ and
$\omega \rightarrow 0$, though it is strongly affected in the vicinity of
$v_{  R}=0$ at low frequencies. In contrast, the component $\chi_{xy}$
(not shown in Fig. 5) stays close to its static value $-(v_{  R}/v_{  D})
\chi_0$ up to the frequency region $\omega \sim \nu$. The real part of the
spin conductivity $\Sigma^z_{yx}$ shows peaks in the vicinity of $v_{  R}=0$
at low frequencies, while the low-frequency behavior of $\Sigma^z_{xy}$
is monotonic in this region, see Fig. 6. According to Eq. (63), the frequency
dispersion of the spin polarizability remains significant at very low
frequencies if $v_{  R}$ is small enough. It is remarkable that in the
"dirty" limit the characteristic rate describing this dispersion,
$(2v_{  R}/v_{  D})^2 \gamma_{  F}$, is equal to $2 (v_{  R}
p_{  F}/\hbar)^2/\nu$, which is the DPK spin relaxation rate for
the Rashba model.

\begin{figure}[ht]
\begin{center}
\includegraphics[scale=0.4]{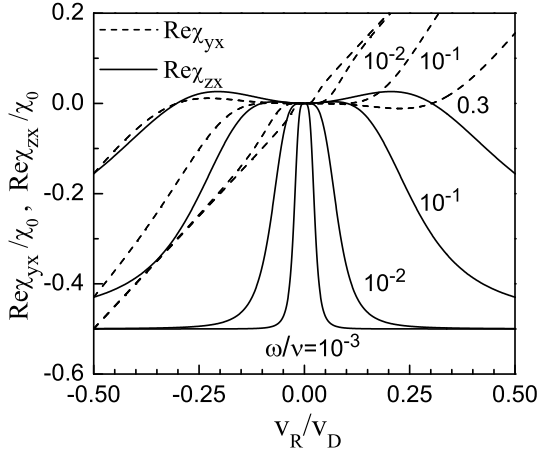}
\end{center}
\addvspace{-0.7 cm}\caption{Dependence of the real part of
spin polarizability in [110]-grown quantum wells on the ratio
$v_{  R}/v_{  D}$ at $v_{  D}p_{  F}/\hbar \nu=1$ for
several frequencies. The polarizability is expressed
in the units of $\chi_0=em v_{  D}/2 \pi \hbar^2 \nu$.}
\end{figure}

\begin{figure}[ht]
\begin{center}
\includegraphics[scale=0.4]{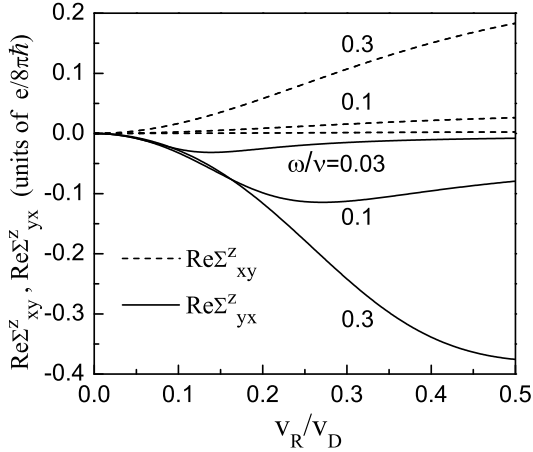}
\end{center}
\addvspace{-0.7 cm}\caption{Dependence of the real part of
spin conductivity in [110]-grown quantum wells on the ratio $v_{  R}/v_{  D}$ at
$v_{  D}p_{  F}/\hbar \nu=1$ for several frequencies.}
\end{figure}

\section{Current response}

It is important to relate the behavior of spin polarization and spin currents
studied in the previous section to the frequency dispersion of the conductivity
(or dielectric function) of 2D electron layers with spin-orbit interaction.
Some relations of this kind have been established previously$^{33,39}$
in the collisionless approximation. In this section the corresponding relations
are obtained and analyzed for the general ${\bf p}$-linear model of spin-orbit
coupling described by the Hamiltonian (1), with taking into account the
electron-impurity interaction. The calculation is based on the
kinetic equation (2). The electric current density is defined as
%64
\begin{equation}
{\bf j}= e \int \frac{d {\bf p}}{(2 \pi \hbar)^2}
{\rm Tr}(\hat{\bf u}({\bf p}) \hat{f}_{\bf p} ).
\end{equation}
Below we neglect the non-parabolicity of the energy spectrum and the
deviation of the Hamiltonian (1) from the linearity, by putting $m_p=m$
and $\xi_p=1$. Let us multiply Eq. (2) by $e{\bf p}/m$, sum it over
${\bf p}$, and take the matrix trace. Using Eqs. (6) and (64), one has
the exact relation
%65
\begin{equation}
-i \omega ( j_{\beta} - 2 e \mu_{\alpha \beta} s_{\alpha}) -
\frac{e^2E_{\beta} n_{  2D}}{m} = e \! \int \! \! \frac{d {\bf p}}{(2 \pi \hbar)^2}
\frac{p_{\beta}}{m} {\rm Tr} \widehat{J}(\hat{f}|{\bf p}).
\end{equation}
The right-hand side of this equation should be set at zero in the collisionless
approximation. To consider the collision-induced contribution in the general case,
one should calculate the distribution function $\hat{f}_{\bf p}$ with the
accuracy up to the terms $\sim (\hbar \Omega_{\bf p} /\overline{\varepsilon})^2$.
Instead of doing this, we consider the limit of short-range scattering potential,
when the right-hand side of Eq. (65) is exactly transformed to $-\nu j_{\beta}$.
Therefore,
%66
\begin{equation}
j_{\beta} = \frac{e^2 n_{  2D}}{m(\nu-i\omega)} E_{\beta} -
\frac{2 i \omega}{\nu-i\omega} e  s_{\alpha} \mu_{\alpha \beta}.
\end{equation}
Equation (66) contains the usual Drude term and the term induced by the
spin-orbit interaction.$^{38,39}$ Together with Eq. (43), this equation
establishes the relationship between the electric current, spin current, and
induced spin density. The validity of both Eq. (43) and Eq. (66) is not restricted
by the assumption of linear response.

Expressing the currents through the
electric conductivity $\sigma_{\beta \lambda}(\omega)$ and spin
conductivity $\Sigma^{\alpha}_{\beta \lambda}(\omega)$, one can write
the equation that relates these tensors:
%67
\begin{eqnarray}
\sigma_{\beta \lambda} = \delta_{\beta \lambda}
\frac{e^2 n_{  2D}}{m(\nu-i\omega)} + \sigma^{  (SO)}_{\beta \lambda}, \nonumber \\
\sigma^{  (SO)}_{\beta \lambda}= \frac{4 m e }{\hbar(\nu-i\omega)}
[{\bmu}_{\beta} \times {\bmu}_{\kappa} ]_{\delta}
\Sigma^{\delta}_{\kappa \lambda}.
\end{eqnarray}
In general, the spin-orbit term $\sigma^{  (SO)}_{\beta \lambda}$
brings non-diagonal contributions to $\sigma_{\beta \lambda}$. This should lead
to a weak Hall effect in the absence of magnetic field at finite frequencies. For the
Rashba model, when only the components $\mu_{xy}=-\mu_{yx}=v_{  R}$ exist, the
conductivity is diagonal and isotropic,$^{47}$ $\sigma^{  (SO)}_{\beta \lambda} =
\sigma^{  (SO)} \delta_{\beta \lambda}$, where
%68
\begin{equation}
\sigma^{  (SO)}= -\frac{4 m e v_{  R}^2}{\hbar(\nu-i\omega)} \Sigma^{z}_{xy}.
\end{equation}
In the collisionless limit, this equation gives a relation$^{39}$ between
the imaginary part of $\sigma^{  (SO)}$ and the real part of the spin-Hall
conductivity $\Sigma^{z}_{xy}$.

In the case of [001]-grown quantum wells with both Rashba and Dresselhaus
types of coupling,
%69
\begin{equation}
\left( \begin{array}{c} \sigma^{  (SO)}_{y \lambda} \\ \sigma^{  (SO)}_{x \lambda}
\end{array} \right)= \frac{4 m e (v_{  R}^2-v_{  D}^2)}{\hbar(\nu-i\omega)}
\left( \begin{array}{c}  - \Sigma^{z}_{x \lambda} \\ \Sigma^{z}_{y \lambda}
\end{array} \right).
\end{equation}
The non-diagonal part of $\sigma^{  (SO)}_{\beta \lambda}$ is related to the
diagonal components of the spin conductivity, while the diagonal part is expressed
through the spin-Hall conductivity (such an expression has been recently
established$^{33}$ in the collisionless regime). The tensor $\sigma^{  (SO)}_{
\beta \lambda}$ can be diagonalized by in-plane rotation of the Cartesian
coordinate axes $OX$ and $OY$. In contrast to the spin polarization, the
quantity $\sigma^{  (SO)}_{\beta \lambda}$ is analytic at $v_{  R}^2=v_{  D}^2$.

In the case of [110]-grown quantum wells the conductivity is diagonal (in the
chosen coordinate system) but anisotropic:
%70
\begin{equation}
\left( \begin{array}{c} \sigma^{  (SO)}_{yy} \\ \sigma^{  (SO)}_{xx}
\end{array} \right)= \frac{4 m e (v_{  R}^2+v_{  D}^2/4)}{\hbar(\nu-i\omega)}
\left( \begin{array}{c}  - \Sigma^{z}_{x y} \\ \Sigma^{z}_{y x}
\end{array} \right).
\end{equation}
The expressions for the components $\Sigma^{z}_{\alpha \beta}$ entering Eqs. (69) and
(70) are obtained from the expressions for spin currents presented in subsections A
and B of the previous section.

Although the contribution $\sigma^{  (SO)}_{\beta \lambda}$ is small as
$(\hbar \Omega_{\bf p} /\overline{\varepsilon})^2$ with respect to the
Drude conductivity, its frequency dependence has qualitatively new features.
If $\Omega_{\bf p} \gg \nu$, so that the spin-split states are well-defined,
the term $\sigma^{  (SO)}_{\beta \lambda}$ describes resonance absorption
of electromagnetic radiation, typically in the THz region,
associated with transitions between these states. In the case of isotropic
spin splitting$^{48}$ (the Rashba model is considered below) and degenerate electron gas,
the resonance takes place$^{38}$ at $\omega \simeq \omega_r \equiv 2 |v_{  R}|
p_{  F}/\hbar$. Substituting in Eq. (68) the detailed expression for the
spin conductivity, see Eq. (46) and Eq. (45) with $\nu_p^{  (1)}=
\nu_p^{  (2)}=\nu$, one has an equation
%71
\begin{equation}
\sigma^{  (SO)}=
\frac{e^2 m v_{  R}^2 i \omega}{\pi \hbar^2 (\nu-i\omega)} ~ \!
\frac{\omega_r^2}{\omega_r^2 \nu - 2i \omega [(\nu-i\omega)^2+\omega_r^2 ]},
\end{equation}
which describes the resonance under consideration at $\omega_r \gg \nu$.
Of course, this resonance also exists in the frequency dependence
of the spin conductivity and spin polarizability.$^{31}$
Another important feature following from Eq. (71) is the presence of
low-frequency dispersion under the opposite condition, $\omega_r \ll \nu$.
This dispersion appears when $\omega$ is comparable to the DPK spin relaxation
rate. Since this rate, for in-plane spin polarization, is given by
$\nu_{  DP}= 2 (v_{  R} p_{  F}/\hbar)^2/\nu$, Eq. (71) in
the limits $\omega_r \ll \nu$ and $\omega \ll \nu$ gives
%72
\begin{equation}
{\rm Re} \sigma^{  (SO)} \simeq - \frac{e^2}{\pi \hbar}
\frac{m v_{  R}^2 \nu_{  DP}}{\hbar \nu^2} \frac{\omega^2}{\omega^2 +
\nu^2_{  DP}}.
\end{equation}
Therefore, $\sigma^{  (SO)}$ essentially depends on $\omega$ in the region
of frequencies $\omega \ll \nu$, when the Drude conductivity still remains
frequency-independent. Another contribution to the frequency dependence of the
conductivity in this region exists owing to weak localization.$^{49}$
Though the weak-localization correction is larger in magnitude
than ${\rm Re} ~ \! \sigma^{  (SO)}$, its frequency dependence is
slow (logarithmic), and can be distinguished from the dependence
given by Eq. (72).

\section{Summary and Discussion}

In this paper, the electric-field-induced spin density ${\bf s}$ and intrinsic
spin current ${\bf q}_{\beta}$ in 2D electron layers described by the general
${\bf p}$-linear spin-orbit interaction Hamiltonian $\hat{\sigma}_{\alpha}
\mu_{\alpha \beta} p_{\beta} \equiv \hat{\bsigma} \cdot {\bmu}_{\beta}
p_{\beta}$ are studied in the classical region of frequencies $\omega$.
The consideration is done for macroscopic systems and at zero magnetic field.
The quantities ${\bf s}$ and ${\bf q}_{\beta}$ are closely related to each
other through Eq. (43) following from the balance equation for spin
density. To find them, a careful analysis of the quantum kinetic equation
is carried out taking into account interaction of electrons with impurities
or other static inhomogeneities. The presented results are valid for
arbitrary correlation between the spin splitting energy
$2 \hbar \Omega_{\bf p}$, disorder-induced broadening $\hbar \nu$,
and energy $\hbar \omega$, under condition that all these energies
are small in comparison with the characteristic kinetic energy of electrons.
A complete analytical solution of the linear-response problem for the matrix
distribution function is given in the limit of short-range scattering potential
(see subsection A of Sec. II). This solution is applied to [001]- and [110]-grown
quantum wells with both Rashba and Dresselhaus types of spin-orbit coupling
(subsections A and B of Sec. III). All other situations when analytical
solutions exist are described is subsections B, C, and D of Sec. II.
The theory also takes into account the isotropic energy-dependent corrections
to the effective mass $m$ (non-parabolicity effect) and to the spin-velocity matrix
$\mu_{\alpha \beta}$. This is reflected by the substitutions $m \rightarrow
m_p$ and $\mu_{\alpha \beta} \rightarrow \xi_p \mu_{\alpha \beta}$ assumed
from the beginning of the consideration. Since these weak corrections do not
lead to qualitative effects (in particular, it is shown that the static spin
currents remain equal to zero in the presence of these corrections), they
are ignored in the most part of the applications, starting from Eq. (42).

The main approximation of the present consideration is the neglect of the
spin-dependent contribution to the scattering potential. This contribution,
also caused by the spin-orbit interaction, is often referred to as the
extrinsic spin-orbit coupling. In the first order with respect to the extrinsic
spin-orbit coupling, there appear spin currents which are not equal to
zero in the static limit. This leads to the extrinsic spin-Hall effect,$^{50,51,52}$
which is beyond the scope of the present paper. The extrinsic spin-orbit
coupling also leads to an additional induced spin polarization,
which is considered, in the limit $\omega \gg \nu \gg \Omega_{\bf p}$,
in Ref. 53. In the second order with respect to the extrinsic spin-orbit
coupling there appear additional relaxation terms in the kinetic equation.
These terms are responsible for the Elliot-Yafet spin relaxation of 2D
electrons.$^{42,43}$ The corresponding spin relaxation rate is many orders
of magnitude smaller that the momentum relaxation rate and often can
be neglected.

The other important approximation is the spatial homogeneity of the problem,
which implies the neglect of the gradient terms in the kinetic equation (2).
For this reason, the results of this paper cannot be directly applied for description
of the spatial distribution of spin density in 100 $\mu$m wide 2D layers
recently studied experimentally.$^{8,28}$ From the theoretical point of view,
the problem of finite-size samples is difficult, not only because of the
presence of gradient terms, but also because of the need to derive the
boundary conditions for the matrix Wigner distribution function. In the general
case, the required boundary condition should be a matrix equation.$^{54}$ Some
simple forms of boundary conditions used in theoretical description of the
spin-Hall effect (see, for example, Ref. 16) have been written without a
derivation. Therefore, the problem of spatially inhomogeneous distribution
of the spin density still remains topical.

The main qualitative result of this study is the prediction of
vanishing spin polarization at arbitrary frequency of the applied electric
field for the special situations when fixed (momentum-independent)
precession axes exist. The well-known examples of these situations
are asymmetric [001]-grown quantum wells with equal Rashba and Dresselhaus
velocities ($v_{  R}^2=v_{  D}^2$) and symmetric [110]-grown quantum
wells ($v_{  R}=0$). This remarkable property, which does not follow
from the consideration of the static response, is explained by the
absence of DPK spin relaxation for these special situations.
As a consequence, in the close vicinity of the fixed precession axis, which
means that $v_{  R}^2$ is close to $v_{  D}^2$ for [001]-grown quantum
wells or $v_{  R}$ is close to zero for [110]-grown quantum wells, the
frequency dispersion of spin polarization is significant in the low-frequency
region, when the polarization can be excited by applying an alternating
electrical bias to the 2D sample. For [110]-grown quantum wells this
dispersion is given by Eq. (63). Since the Rashba velocity is variable by
modifying the shape of the quantum well via a bias applied to an external
(top) gate, these phenomena can be investigated experimentally. As concerns
possible technological use, these phenomena are of interest for the purposes
of frequency filtering and amplification in future spintronics devices.

It should be noted that the predicted behavior in the vicinity of the
fixed precession axes can be influenced by the extrinsic spin-orbit coupling.
From this point of view, the [001]- and [110]-grown quantum wells appear
to be in quite different positions. In [001]-grown quantum wells with
$v_{  R}^2=v_{  D}^2$, where the fixed precession axis lies in the
plane of motion, a coupling of $z$-polarized spins with in-plane
polarized spins occurs when the spin-dependent contribution to the
scattering potential is taken into account. This means that the applied
electric field can excite the spin density perpendicular to the precession
axis as a combined effect of extrinsic spin-Hall current excitation and precession.
Also, as mentioned above, there appears the Elliot-Yafet relaxation
of the in-plane spin density. As a result, finite spin polarizations
will appear for the directions both perpendicular and parallel to the
precession axis, and the non-analyticity with respect to the order of
limiting transitions $v_{  R}^2 \rightarrow v_{  D}^2$ and
$\omega \rightarrow 0$ will be removed. Still, the frequency dispersion
of the spin polarizability should persist down to the frequencies
comparable with the Elliot-Yafet relaxation rate $\nu_{  EY}$.
Using the estimate $\nu_{  EY}/\nu \sim 10^{-5}$ given in
Ref. 42 and taking into account that $\nu \sim 10^{12}$ $s^{-1}$,
one finds $\nu_{  EY} \sim 10^{7}$ $s^{-1}$, which means that the
frequency dispersion can be principally observed in the region above 10 MHz.
On the other hand, in [110]-grown symmetric quantum wells, where the fixed
precession axis is perpendicular to the plane of motion, the applied
electric field cannot excite the in-plane polarized spins even in the
presence of the extrinsic spin-orbit coupling. Excitation of the current
of $z$-polarized spins (the extrinsic spin-Hall current) does not lead
to spin polarization in $z$ direction far from the boundaries of the
sample. Finally, the Elliot-Yafet relaxation of $z$-polarized spin density
is absent in [110]-grown symmetric quantum wells. In conclusion, the
extrinsic spin-orbit coupling does not lead to a finite spin polarization
in [110]-grown symmetric quantum wells and cannot modify Eq. (63).
Modification of the low-frequency behavior described by this equation
may occur owing to non-Markovian memory effects$^{55}$ and effects of
spatial inhomogeneity due to finite sample size.

The other important result is Eq. (66), which establishes a simple relation
between the electric current and induced spin polarization. Though the
validity of this equation is restricted by the approximation of short-range
scattering potential, Eq. (66) is applicable to the whole class of the systems
described by the ${\bf p}$-linear spin-orbit Hamiltonians and does not
require the linear response approximation. It can be useful in applications
and in analysis of experiments where both spin polarization and electric
conductivity are measured. Equation (66) also
shows that the spin-orbit term in the electric conductivity is of purely
dynamic origin, this term vanishes at $\omega \rightarrow 0$. It is not
clear whether this property is specific for the short-range scattering
potential or remains valid for arbitrary scattering potential. The
corresponding calculations are now in progress. In combination with
Eq. (43), the result (66) relates the electric current with the spin current
and leads to a unified description of spin and charge response to the
applied electric field.

The results for frequency dispersion of the spin polarization and spin current
obtained in this paper can be directly reformulated for description of transient
spin response,$^{18,29,31}$ which is investigated experimentally by the
time-resolved spectroscopy.$^{6,11,56}$ If the spin-split electron states are
well-defined, the transient process shows coherent oscillations. If the spin
splitting is suppressed by collisional broadening, there should appear long-time
transients associated with the DPK spin relaxation.$^{18,31}$ Under conditions close
to the appearance of the fixed precession axis, for example, when $v_{  R}$ is
close to zero in [110]-grown quantum wells, the duration of the transient
process, according to Eq. (63), is expected to increase dramatically.

\acknowledgments{
The author is grateful to Sergey Tarasenko for a helpful discussion.}


\begin{thebibliography}{56}
\bibitem{48}
If the conductivity tensor is isotropic, the absorption coefficient
for the electromagnetic wave normally incident on a 2D layer is
equal to $4 \pi {\rm Re} \sigma(\omega)/c\sqrt{\epsilon}$, where
$c$ is the velocity of light and $\epsilon$ is the dielectric
permittivity.
\end{thebibliography}
\end{document}